\documentclass[a4paper,11pt]{article}
\usepackage{pos}
\usepackage{overpic}
\usepackage{natbib}

\title{Prospects on detection of the Fermi Bubbles
with the Cherenkov Telescope Array Observatory}

\author*[a]{F. Xotta}
\author[a]{N. Bavdaž}
\author[a]{C. Eckner}
\author[b]{D. Malyshev}
\author[a]{J. Pérez-Romero}
\author[a]{G. Zaharijaš}

\affiliation[a]{Center for Astrophysics and Cosmology, University of Nova Gorica\\Vipavska 11c, 5270, Ajdovščina, Slovenia }

\affiliation[b]{Friedrich-Alexander-Universit\"at Erlangen-N\"urnberg, Erlangen Centre for Astroparticle Physics\\ Nikolaus-Fiebiger-Str. 2, Erlangen 91058, Germany}

\emailAdd{francesco.xotta@ung.si}

\abstract{
In 2010, the Fermi Gamma-ray Space Telescope observed two gamma-ray emitting structures, the Fermi Bubbles (FBs), which extend up to 55° above and below the Galactic plane and that seem to emanate from the Galactic center region. Although the spectrum at latitudes $|b| > 10^\circ$ has a softening or a cutoff around 100 GeV, the one at the base of the FBs, $|b| <10^\circ$, extends up to $\sim1$ TeV without a significant cutoff in the {\it Fermi}-LAT data. The mechanism behind the FBs production is currently under debate. More observations of the FBs at different energies are required to improve our understanding of their origin.

Recently, H.E.S.S. and HAWC observatory have set upper limits on the FBs. In this work, we assess the sensitivity of the Cherenkov Telescope Array Observatory (CTAO) using the "Alpha configuration" in the South site to detect the FBs and investigate the optimal strategies for their detection at low latitudes. We simulate the observations using the official CTAO science tool gammapy, considering several benchmark models for the FBs and the interstellar emission and test different observational strategies taking advantage of the proposed CTAO consortium surveys. We use these simulations to estimate the CTAO sensitivity to the FBs.
}

\FullConference{High Energy Astrophysics in Southern Africa (HEASA2025)\\
16-20 September, 2025\\
University of Johannesburg, South Africa\\}

\begin{document}
\maketitle

\section{Introduction}
The Fermi Bubbles (FBs) were first identified in 2010 through the observations of the $Fermi$ Large Area Telescope (LAT) \citep{2010ApJ...724.1044S_Su_bubbles}. These large-scale gamma-ray structures extend roughly $55^\circ$ above and below the Galactic plane (GP), originating from the Galactic center (GC). The two lobes are approximately symmetric, though slight asymmetries were observed within the innermost $\sim 6^\circ$. At low latitudes ($|b|<10^\circ$), the emission is brighter and characterized by a hard spectrum with a spectral index of $\sim 2$, whereas at higher latitudes ($|b|>10^\circ$) the intensity decreases and the spectrum exhibits a softening around $\sim 100$ GeV \citep{Fermi}.
Despite targeted searches to observe gamma-ray emission from the FBs with Imaging Atmospheric Cherenkov Telescopes (IACTs), no detection has yet been claimed. The upper limits obtained from H.E.S.S. observations \citep{HESSPaper} suggest a possible cutoff in the low latitude spectrum near $\sim 1$ TeV. The presence of such a cutoff, however, remains highly uncertain. Probing the FBs in the vicinity of the GC at energies around $\sim 1$ TeV is therefore essential to address the existence of a spectral cutoff and provide key insights into their morphology. 

The Cherenkov Telescope Array Observatory (CTAO) represents the next generation of IACTs, designed to operate from 20 GeV to 300 TeV (with peak sensitivity near $\sim 1$ TeV), with an expected increase in sensitivity of one order of magnitude compared to current IACTs%
\footnote{\url{https://www.ctao.org/for-scientists/performance/}}. CTAO will be able to scan the whole sky thanks to its two placements: La Palma (Spain) for CTAO North and the Atacama Desert (Chile) for CTAO South.
Owing to its unprecedented sensitivity and planned surveys of the GC region, CTAO has the potential to detect the FBs emission at high energies, complementing the low energy detection by {\it Fermi}-LAT \cite{Yang_2019}. In this contribution, we evaluate the sensitivity of the CTAO-South array to detect gamma-ray emission from the low latitude FBs.

\section{Methodology}
\begin{figure}
    \centering
    \begin{overpic}[width=0.3\linewidth]{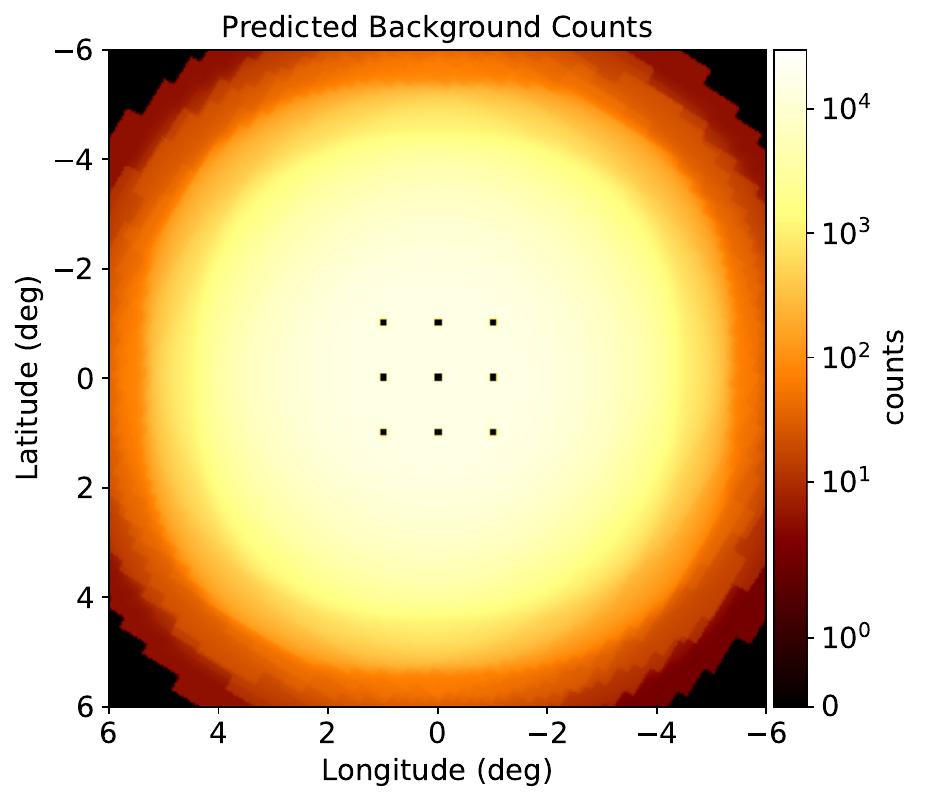}
    \put(15,12){\color{blue} PRELIMINARY}
    \end{overpic}
    \begin{overpic}[width=0.3\linewidth]{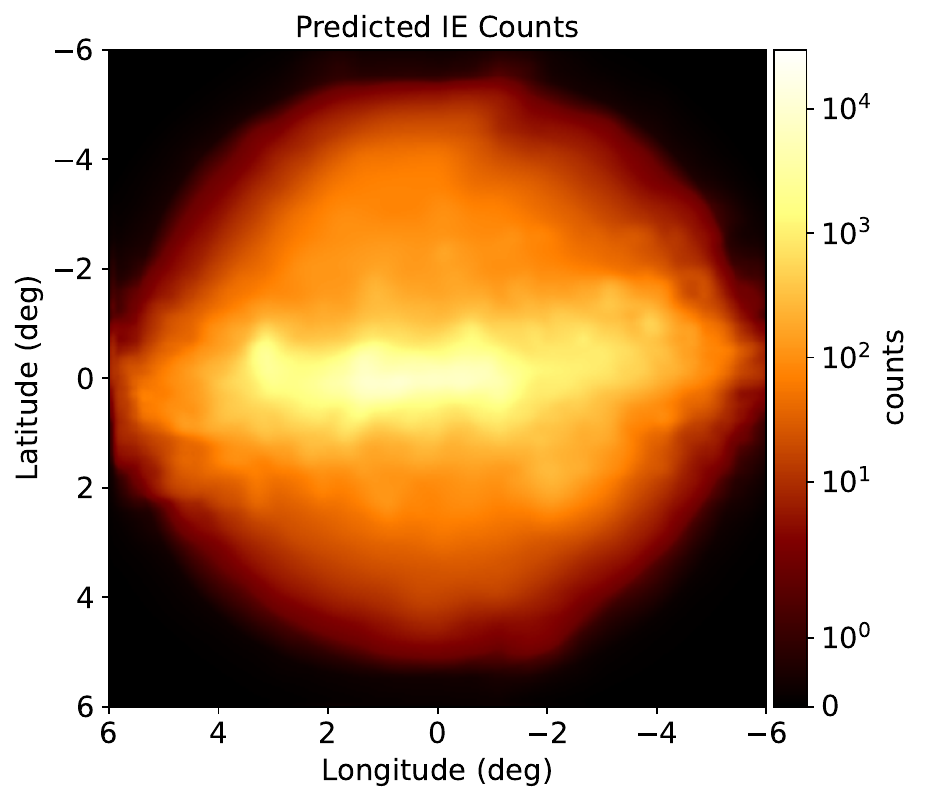}
    \put(15,12){\color{blue} PRELIMINARY}
    \end{overpic}
    \begin{overpic}[width=0.3\linewidth]{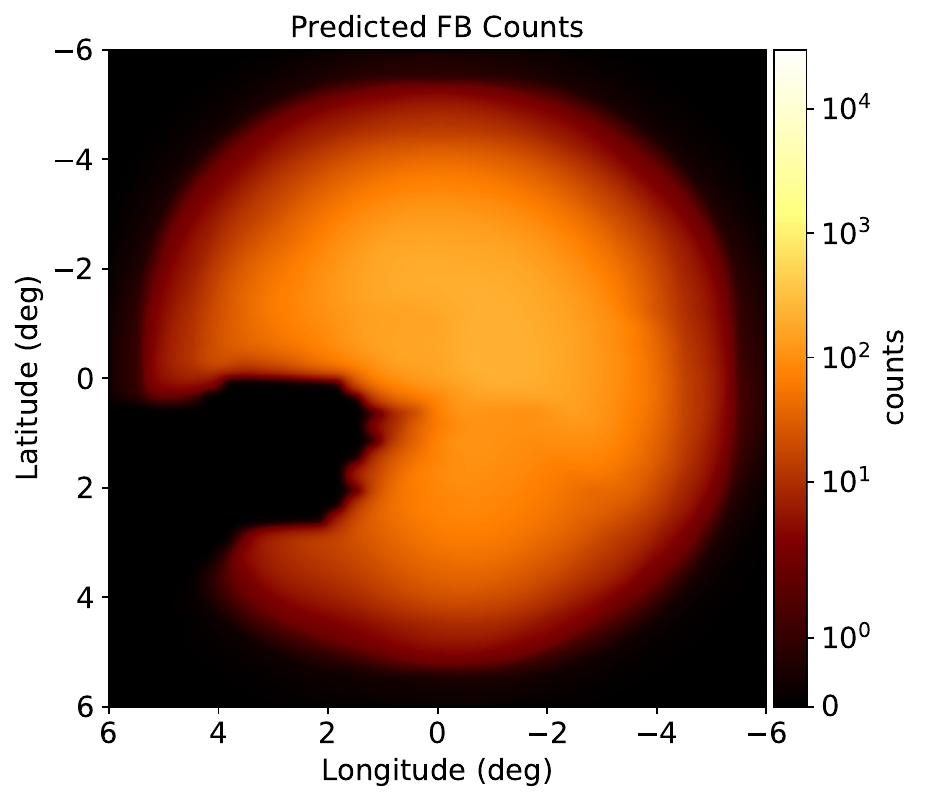}
    \put(15,12){\color{blue} PRELIMINARY}
    \end{overpic}
    \caption{\textbf{Left:} Counts map for CR Background including the pointing positions of the GC survey. \textbf{Center:} Counts map for the Interstellar Emission (IE), corresponding to the VariableMin benchmark model. \textbf{Right:} Counts map for the FBs. All of the plots show the sum of counts over the whole considered energy range.}
    \label{fig:pointingsandmaps}
\end{figure}

\begin{figure}
    \centering
    \begin{overpic}[width=0.5\linewidth]{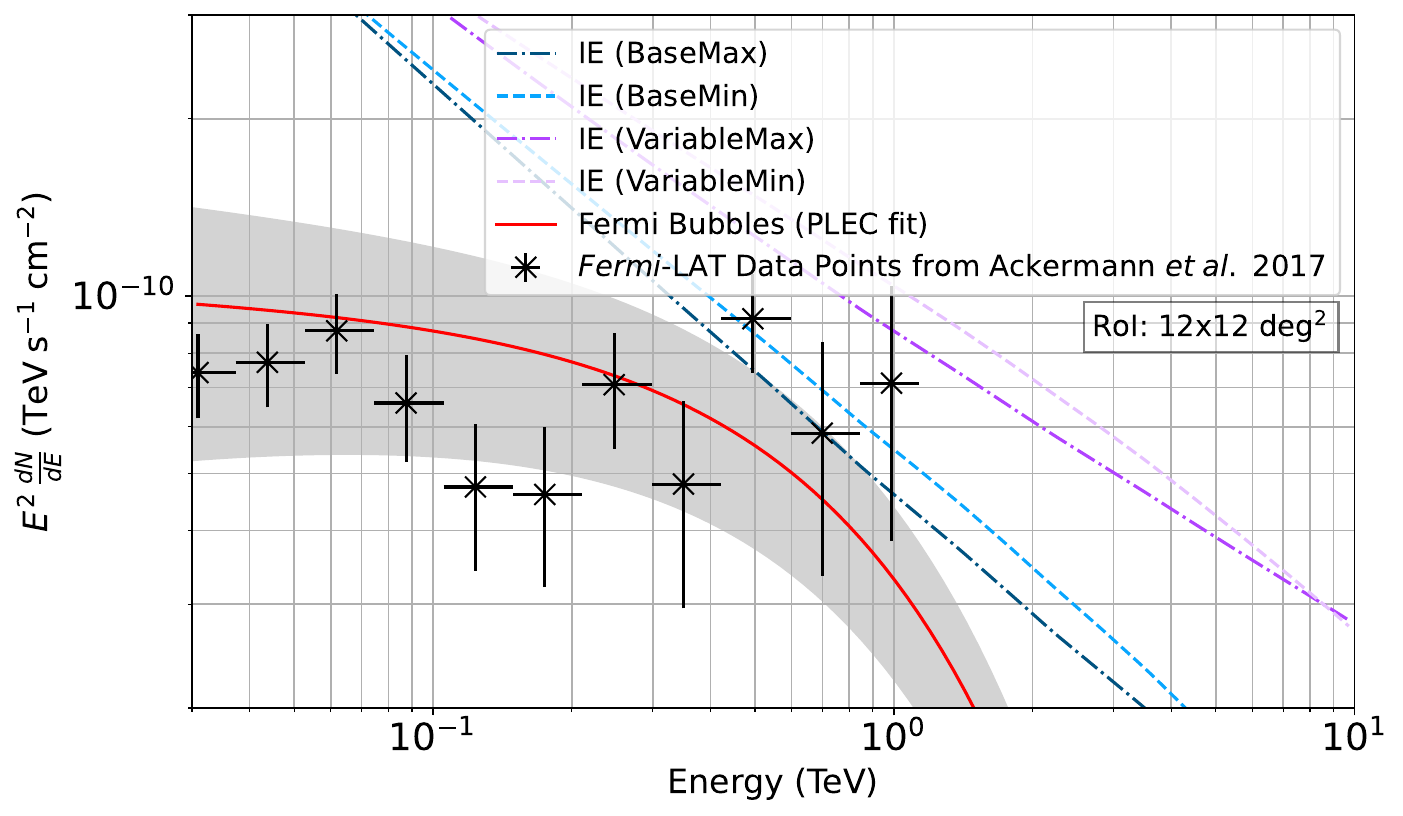}
    \put(20,12){\color{red} PRELIMINARY}
    \end{overpic}
    \caption{Models for the different emission components in our RoI. The FBs model is obtained by fitting the low latitude spectrum in \cite{Fermi} (black stars). 
    We take the Max and Min benchmarks for both the Base and Variable IE models of \cite{DeLaTorre} (4 models in total). 
    }
    \label{fig:IEMComparison}
\end{figure}

\textbf{Simulations.} Our study on the low latitude FBs is based on simulations of observations using the GC survey, consisting of 9 pointings between $b,l=-1,0,1$° with a total time of 525 hours. We define the region of interest (RoI) as 12°$\times$12° centered in the GC (as depicted in the left panel of Figure \ref{fig:pointingsandmaps}).
For this study we consider CTAO South in the "Alpha configuration". We employ the latest instrument response functions (IRFs), "prod5v0.1" \cite{irfs}, which account for this configuration. To assess the detection capabilities of early intermediate CTAO configurations, we simulate the same set-up with a reduced observation time of 50 hours. In this study, the following emission components are included (their counts maps and spectra are shown in Figures \ref{fig:pointingsandmaps} and \ref{fig:IEMComparison} respectively):
\begin{itemize}
    \item {\it Fermi Bubbles:} We use the FBs component from the {\it Fermi}-LAT observations from \cite{Fermi} to derive a model based on a power law with exponential cut-off (PLEC), given as $\frac{dN}{dE}=N_0\left(\frac{E}{1\;{\rm GeV}}\right)^{-\gamma}e^{-E/E_{cut}}$, where $N_0$ is the normalization, $\gamma$ is the index and $E_{cut}$ is the cutoff energy. 
    The fitted parameters are shown in Table \ref{tab:PLECFunc}. 
    \item {\it Interstellar Emission:} The IE, originated from the production of gamma-rays from the interstellar medium in the galaxy, is included using the benchmark models established in \cite{DeLaTorre}. Through these benchmark models we quantify the impact on the uncertainty in the cosmic-ray (CR) diffusion coefficient of the IE for our results.
    \item {\it Instrumental Cosmic Ray Background:} An estimation for the instrumental CR background is taken into account using the model provided by the IRFs, developed within the CTAO Consortium in \cite{irfs}.
\end{itemize}
    \begin{table*}
    \begin{center}
    \begin{tabular}{lccccccc}
    \hline
   \footnotesize Parameter &\footnotesize Value \\
    \hline
    \footnotesize Amplitude, $N_0$, (cm$^{-2}$ s$^{-1}$ TeV$^{-1}$) &\footnotesize $(9\pm2)\times10^{-11}$ \\
   \footnotesize Spectral Index, $\gamma$ &\footnotesize $2.03\pm0.05$ \\
   \footnotesize Cutoff Energy, $E_{cut}$, (TeV) & \footnotesize$1.0$ \\
    \hline
    
    \end{tabular}
    \caption{\label{tab:PLECFunc} Value of the parameters for the PLEC obtained from fitting the low latitude FB data points from \cite{Fermi} with a fixed cutoff energy.}
    \end{center}
    \end{table*}

To generate the mock observations we use Gammapy (v.1.2) \cite{gammapy:2023}, the official CTAO analysis tool, considering 55 logarithmically spaced energy bins between 30 GeV and 100 TeV. We mask the inner $|b|<1.5^\circ$ of the GP to reduce contamination from other emission components (check Figure \ref{fig:pointingsandmaps}).

\textbf{Statistical Framework.} We perform the analysis using a template fitting approach applied to the simulated observations, with the  likelihood based on Poisson statistics for the distribution of the counts.
The predicted counts are defined as $\mu_{ij}(A^{FB}_{i}, A^{IE}_{i}, A^{CR}_{i})=A^{FB}_{i} {\mu^{FB}_{ij}}+A^{IE}_{i} {\mu^{IE}_{ij}}+A^{CR}_{i} {\mu^{CR}_{ij}}$ where the parameters ${{A_i}^{FB,IE,CR}}$ are the normalizations (for each energy bin, labeled as $i$), and the labels are chosen according to the which model they account for (FBs, IE or CR background).\\
Then, we define the test statistics ({\it TS}) as: 
\begin{equation}
    TS({A^{FB}}_i)= 2 {\rm ln}\left[\frac{\mathcal{L}(\mu({A}^{FB}_i, \hat{A}^{IE}_i, \hat{A}^{CR}_i)|n)}{\mathcal{L}(\mu(0,\hat{A}^{IE;0}_i,\hat{A}^{CR;0}_i))|n)} \right],
\end{equation}
where $\hat{A}^{IE,CR}$ represents the norms optimized for a given value of $A^{FB}$ (with $\hat{A}^{IE,CR;0}$ being optimized for $A^{FB}=0$), while $n$ represents the mock (simulated) data. The TS compares hypothesis $H_0$, where $A^{FB}_i=0$, to $H_1$, where $A^{FB}_i\neq0$. These specify that the parameters live respectively in sets $\Theta_0$ and $\Theta_1$. Wilk's theorem states that the TS will then tend to a $\chi^2$ distribution with $\nu=\dim(\Theta_0\cup\Theta_1)-\dim(\Theta_0)$ degrees of freedom. Since $\nu=1$ here, we require a threshold of $TS>4$ for detected points in each bin, roughly equivalent to a 2$\sigma$ detection. 
To compute the mean expected value $\left<A^{FB}_i\right>$, we employ a representative dataset, the Asimov dataset, which corresponds to the expected number of counts in the limit of an infinitely large sample of Poisson realizations.

We further study the impact of correlated systematic uncertainties. To this end, we introduce the scaling parameters $\alpha_{ij}$ and $\beta_i$. $\alpha_{ij}$ represents the differential acceptance uncertainties (which vary through energy and spatial bins), while $\beta_i$ describes the systematic uncertainties for a given energy bin. We assume Gaussian nuisance likelihoods for $\alpha_{ij}$ and $\beta_i$ with respective standard deviations $\sigma_\alpha$ and $\sigma_\beta$\cite{systematics}. The likelihood is then given by:
\begin{equation}
    \mathcal{L}(\mu|n)=\prod_i \frac{1}{\sqrt{2\pi}\sigma_\beta}e^{-\frac{(1-\beta_i)^2}{2{\sigma}^2_\beta}}\prod_{j}\frac{{(\mu_{ij}\alpha_{ij}\beta_i)}^{n_{ij}}}{\sqrt{2\pi}\sigma_\alpha n_{ij}!} \times e^{-\mu_{ij}\alpha_{ij}\beta_i} e^{-\frac{(1-\alpha_{ij})^2}{2{\sigma}^2_\alpha}},
\end{equation}
where:
\begin{equation}
    \alpha_{ij}=\frac{1}{2}\left(1-{\sigma^2_{\alpha}}\mu_{ij}\beta_i+ \sqrt{1+4\sigma^2_{\alpha}n_{ij}-2\sigma^2_\alpha\mu_{ij}\beta_i+\sigma^4_\alpha\mu^2_{ij}\beta^2_i}\right).
\end{equation}
We therefore examine the effect of fixing $\beta_i = 1$ for all $i$, leaving only $\sigma_\alpha$ as a free parameter. 

\textbf{Estimation of the energy cutoff.} Finally, we investigate the capability of CTAO to constrain the energy cutoff of the low latitude FBs emission. We simulate observations fixing the cutoff energy and spectral index of the PLEC to $E_{cut}=1,3,10,30,100,300$ TeV and $\gamma=1.94,2.00,2.10$ respectively, then proceed with the likelihood fit as previously explained. Then, we perform a least $\chi^2$ global fit to the resulting points using a PLEC and a Power Law (PL). Since the only additional parameter in the PLEC with respect to the PL is the cutoff energy, the difference in reduced $\chi^2$ between the two fits should follow a $\chi^2$ distribution with one d.o.f., providing a suitable proxy for the {\it TS} as an estimate of CTAO capability to recover the energy cutoff. 

\section{Results and Conclusions}
\begin{figure}
    \centering
    \begin{overpic}[width=0.4\linewidth]{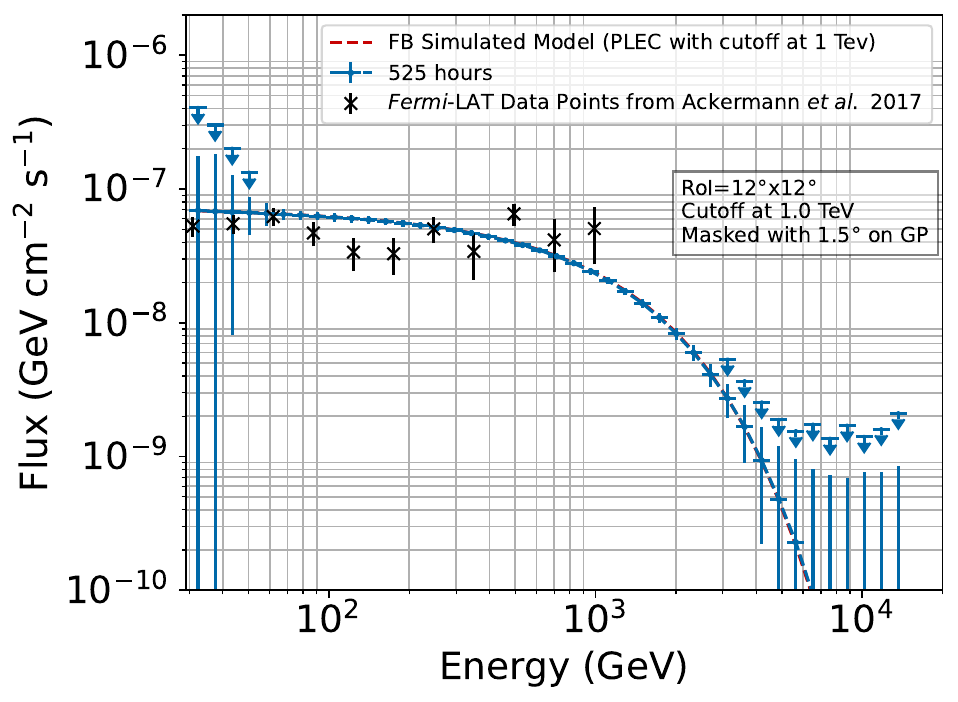}
    \put(20,14){\color{red} PRELIMINARY}
    \end{overpic}
    \begin{overpic}[width=0.38\linewidth]{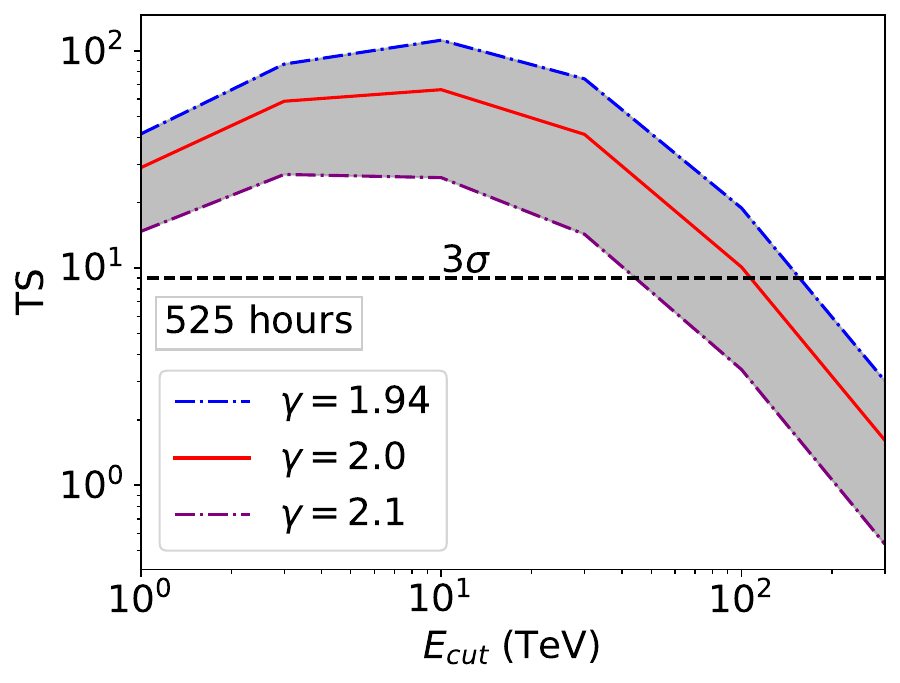}
    \put(50,14){\color{red} PRELIMINARY}
    \end{overpic}
    \caption{\textbf{Left:} Recovery of the injected FBs spectrum for 525 hours in our RoI. Upper limits are shown together with $1\sigma$  uncertainties for each point. \textbf{Right:} Capability of CTAO to recover the energy cutoff for different spectral indices.}
    \label{fig:FBsrecovery}
\end{figure}

\begin{figure}
    \centering
    \begin{overpic}[width=0.4\linewidth]{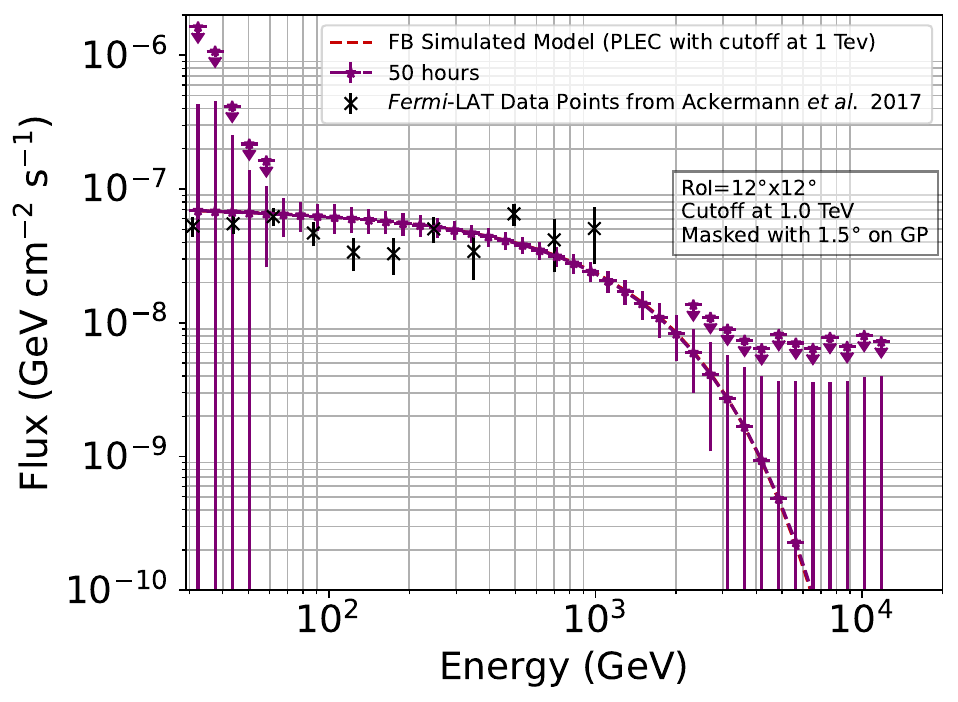}
    \put(20,14){\color{red} PRELIMINARY}
    \end{overpic}
    \caption{Recovery of the injected FBs spectrum for 50 hours in our RoI. Upper limits are shown together with $1\sigma$  uncertainties for each point.}
    \label{fig:FBsrecovery50h}
\end{figure}

\textbf{Recovery of the FBs.} The preliminary results indicate that CTAO should be able to detect the FBs from $\sim60$ GeV to $\sim5$ TeV for a model with a cutoff at 1 TeV assuming no systematic uncertainties. CTAO is further expected to constrain the FBs energy cutoff at the $3\sigma$ level for cutoff values in the range between $\sim 1$ TeV to $\sim 150$ TeV (see the left panel in Figure \ref{fig:FBsrecovery}). For larger spectral indices the PLEC better resembles a PL, lowering the ability of CTAO to differentiate between the two models (check the right panel in Figure \ref{fig:FBsrecovery}).
Reducing the observation time by a factor of 10 still enables a detection from $\sim200$ GeV to $\sim1$ TeV for the same injected model. In this case, the uncertainties grow on average by a factor of $\sim20\%$, as shown in Figure \ref{fig:FBsrecovery50h}.
All the results shown were obtained using the VariableMin benchmark IE model.

\begin{figure}
    \centering
    \begin{overpic}[width=0.405\linewidth]{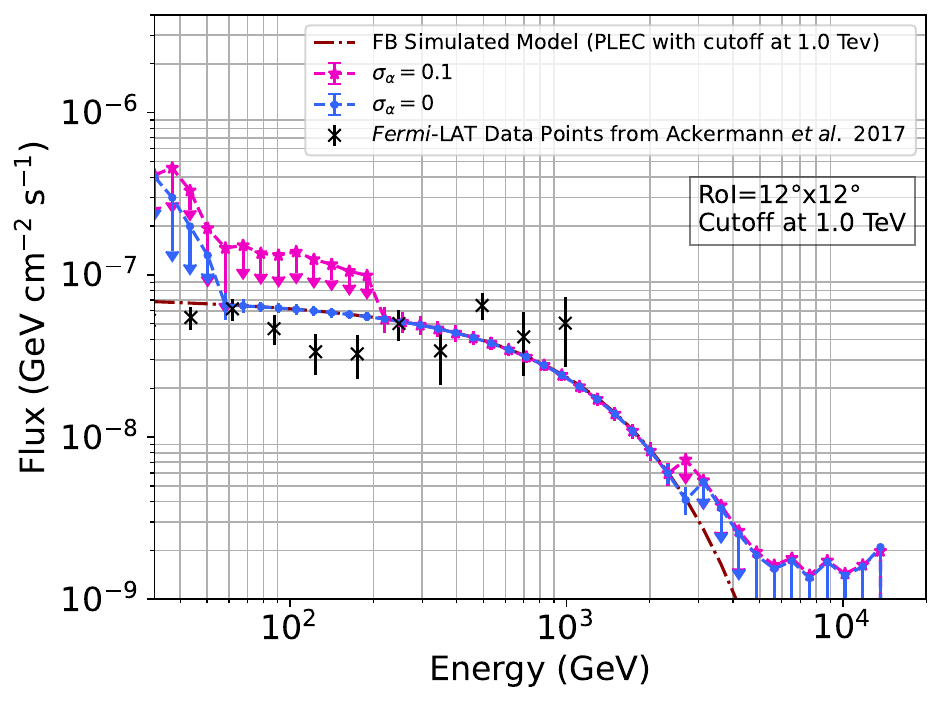}
    \put(20,14){\color{red} PRELIMINARY}
    \end{overpic}
    \begin{overpic}[width=0.4\linewidth]{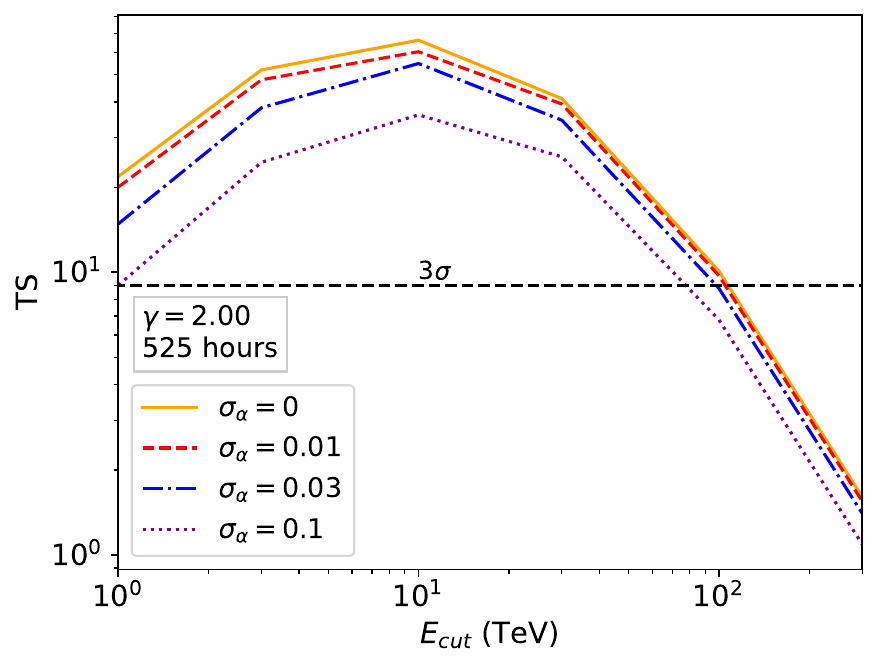}
    \put(45,14){\color{red} PRELIMINARY}
    \end{overpic}
    \caption{\textbf{Left:} Impact of systematic uncertainties on the recovery of the FBs. \textbf{Right:} Impact of systematic uncertainties on the estimation of the energy cutoff.}
    \label{fig:systcompare}
\end{figure}

\textbf{Systematic Uncertainties.} In this study we keep the systematic uncertainties to the benchmark values $\sigma_\alpha=1\%,3\%,10\%$. This impacts our results as an average decrease of $4\%,36\%,77\%$ in the {\it TS}, respectively. In particular, for $\sigma_\alpha = 10\%$, the FBs model is recovered above ${\sim}200$ GeV, with an associated increase up to $\sim30\%$ in the flux uncertainty (as seen in the left panel in Figure \ref{fig:systcompare}). The impact on the cutoff estimation is comparatively modest, degrading the {\it TS} by up to $\sim 10\%$, as can be seen in the right panel of Figure \ref{fig:systcompare}.

\textbf{Summary and future prospects.} Our results show that the CTAO is expected to succesfully detect the emission from the FBs. The FBs can be recovered from ${\sim}60$ GeV without systematic uncertainties, and from ${\sim}200$ GeV for $\sigma_\alpha = 10\%$ for an injected model with a PLEC with a cutoff at 1 TeV. \\
Future work will focus on assessing in greater detail the prospects for an early detection of the FBs and the characterization of a potential spectral cutoff. We will also investigate the sensitivity of CTAO subarrays, and extend our analysis to the high latitude FBs.

\acknowledgments
Francesco Xotta's contribution is supported by the the Slovenian Young Researcher Program. Christopher Eckner's and Judit Pérez-Romero's contributions are supported by the European Union's Horizon Europe research and innovation programme under the Marie Skłodowska-Curie Postdoctoral Fellowship Programme, SMASH co-funded under the grant agreement No. 101081355.

\bibliographystyle{JHEP}
\bibliography{bibliography}

\end{document}